# Wave amplification in the framework of forced nonlinear Schrödinger equation: the rogue wave context

Alexey Slunyaev[1,2),*], Anna Sergeeva[1,2)] and Efim Pelinovsky[1,2)]

[1)] Institute of Applied Physics, N. Novgorod, Russia
[2)] Nizhny Novgorod State Technical University, N. Novgorod, Russia

**Abstract**

Irregular waves which experience the time-limited external forcing within the framework of the nonlinear Schrödinger (NLS) equation are studied numerically. It is shown that the adiabatically slow pumping (the time scale of forcing is much longer than the nonlinear time scale) results in selective enhancement of the solitary part of the wave ensemble. The slow forcing provides eventually wider wavenumber spectra, larger values of kurtosis and higher probability of large waves. In the opposite case of rapid forcing the nonlinear waves readjust passing through the stage of fast surges of statistical characteristics. Single forced envelope solitons are considered with the purpose to better identify the role of coherent wave groups. An approximate description on the basis of solutions of the integrable NLS equation is provided. Applicability of the Benjamin – Feir Index to forecasting of conditions favourable for rogue waves is discussed.

## 1. Introduction

In this paper we consider the evolution of ensembles of irregular nonlinear waves, and also the dynamics of envelope solitons subjected to external forcing. This problem appears in different realms and has a general physical significance. In the present study we focus on the most simplified setting, when narrow-banded waves propagate along one coordinate; the evolution equation includes terms of weak linear dispersion and nonlinear four-wave interactions. The evolution of the conservative system is governed by the nonlinear Schrödinger (NLS) equation of focusing type, which supports existence of localized stable wave groups – 'bright' envelope solitons; it is known to be integrable by means of the Inverse Scattering Technique after Zakharov & Shabat (1972). The external action on the wave system is taken into account through an additional term of linear forcing/damping, which in dimensionless form is of the order of magnitude similar to the squared wave steepness. This modification violates integrability of the NLS equation, thus approximate or numerical approaches become preferable.

So-called rogue waves are a particular case when the nonlinear Schrödinger theory sheds light on the physical mechanisms, responsible for a sudden occurrence of unexpectedly high waves. Rogue waves form a specific class of extreme waves which occur much more frequent than it is prescribed by the statistics of the random Gaussian process. The role of the modulational (Benjamin – Feir) instability in generation of rogue waves on the sea surface is generally accepted at present (see reviews Kharif & Pelinovsky, 2003; Dysthe et al, 2008;

---

[*] slunyaev@hydro.appl.sci-nnov.ru

Kharif et al, 2009; Slunyaev et al, 2011). A recent survey of rogue wave phenomena in various media supporting nonlinear wave self-modulation may be found in Onorato et al (2013). The theory for modulational instability of wind driven plane waves which is described by the forced NLS equation was developed by Leblanc (2007), though the opposite problem of stability of weakly damped waves was discussed in Segur et al (2005). The Benjamin-Feir instability of uniform waves may become explosive in the case of external pumping. In the context of water waves the NLS equation with forcing term was discussed in paper by Onorato & Proment (2012) (see also references therein). More general extensions of the nonlinear Schrödinger equation in application to the wind driven surface waves and possible mechanisms of wave growth saturation were considered by Fabrikant (1980). The theory under assumption of stronger wind, which is of the same order as wave steepness, was developed by Brunetti et al (2014). New terms which appear in that modification of the NLS equation may be eliminated after appropriate change of variables, and then the equation turns to the form considered in (Leblanc, 2007; Kharif et al, 2010; Onorato & Proment, 2012).

Numerical simulations of individual wave groups affected by wind were performed within different frameworks in (Yan and Ma, 2010; Adcock and Taylor, 2011). Laboratory and numerical simulations of transient and modulated wave trains in the presence of wind were conducted by Touboul et al (2006), Kharif et al (2008, 2010), Chabchoub et al (2013a).

It is important that most of statistical theories are based on assumptions of stationarity and proximity to gaussianity. Meanwhile, strongly non-stationary situations seem prone to higher probability of extreme waves, such as: the transition from non-equilibrium condition to the quasi-stationary one (Janssen, 2003; Socquet-Juglard et al, 2005; Onorato et al, 2009; Annenkov & Shrira, 2009a; Shemer et al, 2010b and references therein), rapidly changing winds (Annenkov & Shrira, 2009a,b), change of the waveguide properties (Janssen & Herbers, 2009; Onorato et al, 2011; Sergeeva et al, 2011; Ruban, 2012; Trulsen & Zeng, 2012). The problem of statistical description of these situations is still challenging.

Higher probability of extreme waves in random fields relates to large values of the fourth statistical cumulant, kurtosis, which may be split into two parts: the dynamical kurtosis due to resonant and near-resonant interactions and the kurtosis due to non-resonant interactions (bound waves) (Mori & Janssen, 2006; Annenkov & Shrira, 2009a). The transition regimes of intense irregular waves observed e.g. by Annenkov & Shrira (2009a), Shemer et al (2010b) were accompanied by an increase of the dynamical kurtosis, which constituted the main part of total kurtosis. In contrast, in typical for sea states quasi-stationary conditions the kurtosis due to bound waves seems to prevail (Annenkov & Shrira, 2009a). In (Slunyaev, 2010; Slunyaev & Sergeeva, 2011) the increase of dynamical kurtosis was associated with formation of soliton-like wave patters due to coherence in wave harmonics, which obviously breaks the assumption of random wave phases.

In this paper we consider how time-limited pumping changes ensembles of irregular nonlinear waves and in particular their statistics. Since the concerned problem is complicated, the role of constrained by nonlinearity wave groups (envelope solitons) is studied in parallel with the help of supplementary numerical simulations. We consider situations of different ratios between the two characteristic time scales which may be singled out: the effective time of nonlinearity and the characteristic time of energy input. In Sec. 2 the problem statement is given, and the key controlling parameters are formulated. In Sec. 3 the results of simulations of irregular waves are reported and discussed from the general viewpoint. They are interpreted later with the help of auxiliary numerical experiments described in Sec. 4, where deterministic envelope solitons are simulated. The solution of the Cauchy problem for the integrable NLS equation in terms of the Inverse Scattering Technique, and the asymptotic solution of the perturbed NLS equation yield qualitative understanding of the observed nonlinear phenomena; they provide with approximate formulas. We discuss the main conclusions including the relation to rogue waves and their forecasting at the end.

## 2. Problem statement and controlling parameters

The framework adopted in this study is the modified nonlinear Schrödinger equation, which may be also considered as a particular case of the complex Ginzburg – Landau equation or more generalized equations for wind driven waves (Fabrikant, 1980, 1984; Brunetti et al, 2014). For the situation of gravity waves over deep water the equation reads

$$i\frac{\partial A}{\partial t} + \frac{\omega_0}{8k_0^2}\frac{\partial^2 A}{\partial x^2} + \frac{\omega_0 k_0^2}{2}|A|^2 A - i\Gamma A = 0, \qquad (1)$$

where the last term is responsible for damping/pumping. In what follows we assume that the system is conservative ($\Gamma = 0$) or experiences energy input ($\Gamma > 0$). The pumping is due to Miles' mechanism of wave generation as derived by Leblanc (2007) (see also Fabrikant (1980)). Function $A(x, t)$ is the complex envelope amplitude; $k_0$ and $\omega_0$ are respectively the wavenumber and frequency of the carrier ($\omega_0^2 = gk$, where, $g$ is the gravity acceleration); the equation is written in co-moving references. The surface displacement $\eta(x, t)$ is determined by relation

$$\eta = \mathrm{Re}(A\exp(i\omega_0 t - k_0 x)), \qquad (2)$$

where the bound wave components are completely ignored in this paper for simplicity of interpretation of results.

For qualitative comprehension of the nonlinear wave dynamics governed by (1) it is instructive to consider a dimensionless representation, when variables are properly scaled. Let us assume that the envelope amplitude $A(x, t)$ is characterized by a typical value $a$ and a typical length, $(\Delta k)^{-1}$ ($\Delta k$ has the meaning of the wavenumber spectrum width). Then it is straightforward to transform Eq. (1) to the following scaled dimensionless form

$$i\frac{\partial A'}{\partial t'} + \frac{1}{BFI^2}\frac{\partial^2 A'}{\partial x'^2} + |A'|^2 A' - irA' = 0, \qquad (3)$$

after changes

$$t' = \frac{t}{T_{nl}}, \qquad x' = \Delta k x, \qquad A' = \frac{A}{a}, \qquad (4)$$

$$T_{nl} = \frac{2}{\omega_0 (k_0 a)^2}, \qquad BFI = 2\frac{k_0^2 a}{\Delta k}, \qquad r = \Gamma T_{nl}.$$

Here $T_{nl}$ is the characteristic time of nonlinear effects. The similarity parameter $BFI$ is the so-called Benjamin – Feir Index. It characterizes the ratio of the nonlinear term over the dispersive term. Value $\omega_0 T_{nl}$ should be large to provide applicability of the weakly-nonlinear theory. Parameter $r$ naturally appears in (3); it is the ratio of the characteristic nonlinear time versus the characteristic time of the forcing (which is proportional to $\Gamma^{-1}$). This parameter controls the significance of the pumping term in comparison to the nonlinear term. To provide the smallness of forcing with respect to the wave advection terms, inequality $\Gamma/\omega_0 \ll 1$ should take place. Note that scaling (4) and consequent equation (3) explicitly embody the scales of the particular wave state under consideration.

Let us specify constant $a$ as the root mean square of the modulus of complex function $|A|$. Note that then in the case of sufficiently long modulations value $a^2$ is about twice the mean square of the displacement, $a^2 \approx 2\eta_{rms}$. Therefore in (4) we may replace $k_0 a$ with $2^{1/2}\varepsilon$, where $\varepsilon = k_0\eta_{rms}$ is the average wave steepness; we also introduce the dimensionless spectrum width $\nu \equiv \Delta k/k_0$. Hence the set of controlling parameters may be defined in the following form

$$T_{nl} = \frac{1}{\omega_0 \varepsilon^2}, \qquad BFI = 2\sqrt{2}\frac{\varepsilon}{\nu}, \qquad r = \Gamma T_{nl}. \qquad (5)$$

The definition of *BFI* in (5) is in accordance with one given in Janssen (2003); irregular waves are modulationally unstable if $BFI > 1$. Coefficients *BFI* and $r$ in equation (3) are the similarity parameters which control the wave evolution. The transformation from (1) to (3) is convenient for the qualitative analysis if the wave field is reasonably well characterized by the wave steepness $\varepsilon$ and the spectrum width $\nu$; these quantities enter the controlling parameters.

In this study the initial condition for numerical simulations of ensembles of irregular waves is characterized by the power spectrum with Gaussian shape. At $t = 0$ the Fourier spectrum of complex amplitudes $A(x, t = 0)$ is the same for all realizations, with uniformly distributed random phases (similar to e.g. Shemer et al, 2010b). According to what is said above, the significant parameters of the initial condition are the steepness $\varepsilon = k_0\eta_{rms}$ and the spectrum width, $\nu \equiv \Delta k/k_0$; they determine *BFI*. We use $\nu(t = 0) = 0.2$ in all simulations of irregular waves reported below. The value $\nu = 0.2$ corresponds to a relatively narrow spectrum, which seems to be feasible in the real sea. For example, in laboratory simulations of wave groups by Shemer at al (2010a) the spectrum with $\nu = 0.2$ corresponded to the most broad-band case, which was characterized by the JONSWAP spectrum with large peakedness $\gamma = 7$. Different $\eta_{rms}(t = 0)$ are used in the present simulations, which result in different steepnesses $\varepsilon$ ($k_0$ is kept the same in all the simulations). Due to the scaling transformation invariance of the NLS equation, only the ratio of $\varepsilon$ over $\nu$ matters (see (5)): waves with narrower spectrum and smaller steepness will result in a similar solution. In different simulation series we chose combinations of $\varepsilon$ and $\nu$ to have values of *BFI* less, about and larger than one.

The spatial domain of simulations is about 60 wave lengths; periodic boundary conditions are applied. In each experimental series we have 200 random wave realizations. Though the simulations were performed for dimensional variables, the results are presented in a dimensionless form to allow general consideration. A typical picture of the initial surface elevation may be found in Section 3.2 (Fig. 5a).

The following problem statement is used in numerical experiments, which models a time-limited action of wind (or another sort of the external forcing). All the experiments begin at $t = 0$, the first stage of evolution is conservative. The energy input starts at $t = T_{on}$ and stops at $t = T_{off}$. The experiment is continued with $\Gamma = 0$ after $T_{off}$ until moment $T_{exp} > T_{off}$.

Quantity $T_{nl}$ (see Eq. (5)) estimates the characteristic time of four-wave nonlinear interactions. It was shown in Slunyaev & Sergeeva (2011) that irregular waves return to quasi-stationary conditions within 1-2 $T_{nl}$ in a wide range of parameters. Simulations performed during this research consist of three stages, providing time-limited forcing: 1) establishing a quasi stationary state during period from 0 to $T_{on}$; 2) pumping energy to the system from $T_{on}$ to $T_{off}$; and 3) establishing a new quasi stationary state during from $T_{off}$ to $T_{exp}$. To provide sufficient time for the system settlement, we choose $T_{on} = 5T^{(0)}_{nl}$ and $T_{exp} = T_{off} + 10T^{(1)}_{nl}$. Note the distinction between two characteristic times $T^{(0)}_{nl}$ and $T^{(1)}_{nl}$ which are computed from (5) for the initial steepness $\varepsilon(0)$ and its increased value $\varepsilon(T_{off})$ respectively ($T_{nl}^{(1)} \leq T_{nl}^{(0)}$).

During the wave evolution the carrier wavenumber $k_0$ remains constant within the NLS equation framework. When forcing is absent ($\Gamma = 0$) the integral of $|A|^2$ over the simulated domain is preserved. Strictly speaking, quantity $\varepsilon$ is not conserved by equation (1), because $\varepsilon$ is specified through the real part of the complex amplitude $A$. Function $\varepsilon(t)$ oscillates with time, though the amplitude of oscillations is small when modulations are long, and may be neglected in all reported experiments. When the pumping term is not zero, the total energy of the system grows according to the exponential law

$$E(t) = E(0)\exp(2\Gamma t), \quad \text{where} \quad E(t) = \int |A(x,t)|^2 dx. \qquad (6)$$

Respectively, steepness $\varepsilon(t)$ grows exponentially with increment $\Gamma$ (in the scaled variables $\varepsilon'(t')$ grows with increment $r$).

Non-dimensional parameter $r$ specifies the regime of wave pumping and is different in various series of experiments. The forcing effect on the wave system is characterized by the duration $T_w = T_{off} - T_{on}$ and by the cumulative amplification of the characteristic wave amplitude, specified by a new parameter $G$. Then coefficient $r$ may be determined through these natural parameters by means of the following formulas,

$$G = \sqrt{\frac{E(T_{off})}{E(0)}} \approx \frac{\varepsilon(T_{off})}{\varepsilon(0)}, \qquad r = \frac{T_{nl}}{T_w} \ln G. \qquad (7)$$

These relations are used to obtain $T_w$ (and then $T_{off}$) for given $\varepsilon$, $G$ and $r$.

**3. Numerical simulations of irregular waves in the forced NLS equation**

*3.1. No pumping (conservative system)*

There are numerous studies both in numerical and laboratory frameworks dealing with investigations of irregular wave evolution in conservative (or with negligible dissipation) systems (among others, Onorato et al, 2001, 2002, 2009; Janssen, 2003; Socquet-Juglard et al, 2005; Gramstad & Trulsen, 2007; Shemer & Sergeeva, 2009; Shemer et al, 2010a,b; Mori et al, 2011). The general picture of this problem when waves propagate in one direction was depicted in Slunyaev & Sergeeva (2011). The initial condition in the form of a random superposition of sinusoidal waves or superposition of Stokes waves leads to essentially non-stationary transition to a new equilibrium state during a couple of characteristic 'nonlinear' times $T_{nl}$. During this transition period the wave spectrum broadens noticeably, while the portion of abnormally high waves increases drastically; the degree of the spectrum transformation is controlled by the Benjamin – Feir Index.

In the first series of our numerical experiments the pumping term is put equal to zero ($\Gamma = 0$ for $0 \leq t \leq T_{exp}$). Other parameters of experiments are given in Table 1. Different regimes of irregular wave evolution are simulated in experiments A1, A2 and A3 for three different values of steepness of the initial conditions. The total duration of experiments A1-3 was limited by fifteen 'nonlinear' times, $T_{exp} = 15T_{nl}^{(0)}$. A sharp stage of the wave system adjustment is observed when $BFI(0) > 1$ (Fig. 1).

To describe the current state of the wave system, its average spectrum is followed up (width and shape), and also the temporal dependence of momentary averaged *BFI* and statistical moments skewness and kurtosis, specified as

$$Sk(t) = \left\{ \frac{\langle (\eta - \bar{\eta})^3 \rangle}{\langle (\eta - \bar{\eta})^2 \rangle^{3/2}} \right\}, \quad Ku(t) = \left\{ \frac{\langle (\eta - \bar{\eta})^4 \rangle}{\langle (\eta - \bar{\eta})^2 \rangle^2} \right\} - 3, \quad \bar{\eta}(t) = \langle \eta \rangle, \quad (8)$$

are tracked. Here angle brackets mean averaging over coordinate for each realization, while curved brackets denote ensemble averaging. The Gaussian random process is characterized by $Sk = 0$, $Ku = 0$. To reduce the effect of heavy spectral tails on the estimation of the spectrum width $\nu$, it is computed on the basis of the power wavenumber spectrum $S(k)$ in the interval of wavenumbers from zero to the second harmonic

$$\nu(t) = \left\{ \frac{1}{k_0} \sqrt{ \frac{\int_0^{2k_0} (k - k_0)^2 S dk}{\int_0^{2k_0} S dk} } \right\}. \quad (9)$$

Table 1. Parameters of numerical experiments. Cases B1 and S1 correspond to very slow pumping.

| | Experiment code | $\varepsilon(0)$ | G | $\varepsilon(T_{off})$ | r | $\Gamma / \omega_0$ |
|---|---|---|---|---|---|---|
| No forcing | A1 | 0.04 | – | – | – | – |
| | A2 | 0.08 | – | – | – | – |
| | A3 | 0.16 | – | – | – | – |
| Forcing of irregular waves | B1 | 0.08 | 2 | 0.16 | 1/5 | 0.0013 |
| | B2 | 0.08 | 2 | 0.16 | 1 | 0.0065 |
| | B3 | 0.08 | 2 | 0.16 | 5 | 0.0324 |
| Forcing of envelope soliton | S1 | 0.08 | 2 | 0.16 | 1/5 | 0.0013 |
| | S2 | 0.08 | 2 | 0.16 | 1 | 0.0065 |
| | S3 | 0.08 | 2 | 0.16 | 5 | 0.0324 |

The integration over all $k \geq 0$ may in fact result in rather different values of $\nu$, hence the spectrum width and the value of *BFI* are not well-defined in some simulations (see in Discussion).

The momentary dependences of *BFI* are calculated for each realization with subsequent ensemble averaging. The confidence intervals, determined as the standard deviation of *BFI*s for given time for different realizations, are shown in Fig. 1 by thin lines of corresponding colours. Confidence intervals for other parameters are calculated similarly. The average steepness $\varepsilon$ is approximately conserved in experiments A1-3 ($E(t) = $ Const).

In experiment A1 the initial value $BFI(0) \approx 0.56$ is less than one, and remains about the same throughout the simulation, see Fig. 1a (note scaled time axes in Fig. 1 and further). Correspondingly, no noticeable change of the spectrum width is observed in series A1. The kurtosis attains small positive value of about 0.5 (Fig. 1b). Steeper wave conditions A2 and A3 correspond to $BFI(0) \approx 1.1$ and $BFI(0) \approx 2.2$ correspondingly, what results in spectral broadening, and thus decreasing of the momentary *BFI* till about one (Fig 1a). The parameter of kurtosis rapidly grows at the very beginning of the simulations (in particular in case A3), and then somewhat restores to smaller values (Fig. 1b). The skewness is small and oscillates about zero in all the cases; this could be expected since the bound waves are disregarded in the reconstruction formula (2), and the NLS equation possesses the phase symmetry.

The amplitude exceedence probability distribution functions are shown in Fig. 1c for all three cases (with confidence intervals) and compared with the Rayleigh distribution valid for the Gaussian random process. The distribution functions are calculated for data within time slots $(T_{off} + T_{exp})/2 < t < T_{exp}$, thus they correspond to the established quasi-stationary state. It is clear that larger initial *BFI* result in higher probability of large waves, what correlates with values of kurtosis.

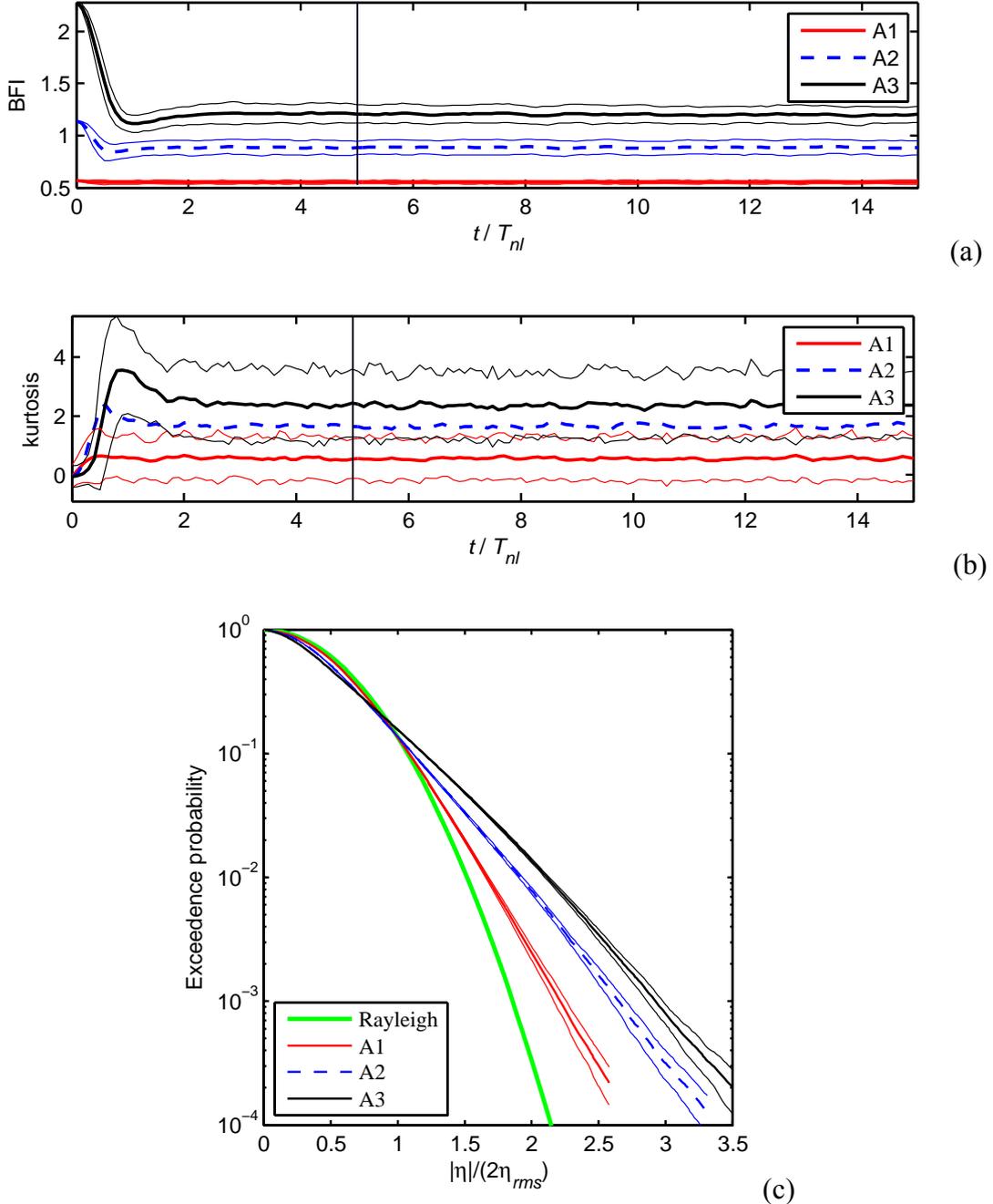

Fig. 1. Evolution of the *BFI* (a) and kurtosis (b) in three cases A1-3. The vertical line marks the instant $5T_{nl}$; the total duration is $15T_{nl}$. The probability distribution functions for the quasi-stationary states compared with the Rayleigh distribution (c). Thin lines of corresponding

colours show the confidence intervals (plus-minus standard deviation); the confidence interval for case A2 is not shown in panel (b).

The detailed description of similar experiments in water wave flumes, and also comparison with the frameworks of the NLS and generalized NLS equations may be found in Shemer et al (2010b). Surprisingly the NLS equation captures the key features of the wave system evolution rather well even when steep near-breaking waves are considered. The results of various numerical and laboratory simulations may be well parameterized in terms of the *BFI* and scaled by $T_{nl}$ time (Slunyaev & Sergeeva, 2011), hence the four-wave interactions govern the wave evolution for the most part.

The attained equilibrium state at $t >> T_{nl}$ in case A3 does not correspond to the Gaussian random process, the value of kurtosis is definitively not small (Fig. 1b) (according to distributions in Fig. 1c, all cases deviate from the Gaussian statistics, but A3 does to a greater extent). Note that in this work we discuss only so-called 'dynamical' kurtosis (bound waves are not taken into account at all). While the kurtosis due to bound waves is always small (of order of $\varepsilon^2$) the dynamical kurtosis may be of order of one (Mori & Janssen, 2006). Positive values of *Ku* reveal formation of nonlinear long-living groups, see discussion in Slunyaev & Sergeeva (2011). Statistical theories usually imply weak non-Gaussianity of random processes, hence they inherently assume that kurtosis *Ku* should not deviate from zero significantly. Case A3 definitively falls out these assumptions.

*3.2. Time-limited pumping*

In the next sequence of numerical experiments (labeled with letter B in Table 1) irregular waves experience time-limited forcing during the period $T_{on} < t < T_{off}$. All the experiments start from the most interesting case of rather intense waves (with initial *BFI* slightly exceeding one, hence the effects of dispersion do not prevail over nonlinear effects) and experience overall two times wave height amplification ($G = 2$). Experiments B1–3 differ in ratios *r* between the characteristic scale of forcing and the nonlinear scale: the instantaneous forcing becomes stronger in cases from B1 to B3. Case B3 corresponds to the rapid pumping, though B1 is the opposite case of a very slow energy input. However, the final states are characterized by the same level of energy. Note that parameter $\Gamma/\omega_0$ is small in all the simulations (see Table 1) thus the forcing is always slow comparing with wave period.

As before, the initial condition for surface elevation is specified as a linear superposition of random sinusoids composing the Gaussian spectrum (the dash-dotted line in Fig. 2). After the start of simulations the wave system attains the quasi-stationary state before the moment when the pumping is switched on. This state is characterized by a slightly different spectral shape compared to the initially prescribed, see thin green line in Fig. 2 (coding 'initial state'). The pumping disbalances the system, while it tends to return to the new equilibrium state. During the stage of forcing a competition between effects of pumping and nonlinear restoration may be intuitively expected, which may be characterized by the single parameter *r*.

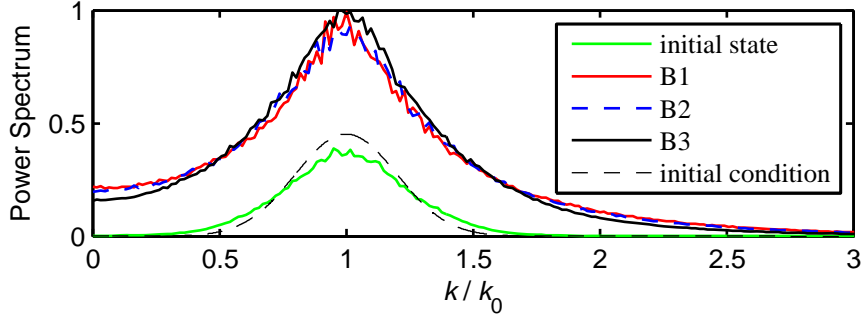

Fig. 2. Wavenumber power spectra of the surface elevation. Thin dashed line shows the initial condition (the Gaussian shape), green solid curve corresponds to the average spectrum before the forcing starts ('initial state'). Other lines show the averaged spectra in the quasi-stationary states of simulations B1-3 after the pumping is turned off.

The time-limited pumping results in increase of the *BFI*, though its time dependence is rather different in different series of simulations, see Fig. 3a. Note that time is scaled over the value $T_{nl}^{(0)}$, which is the same for all cases the B1–3. When the forcing is short in time and hence effectively stronger (large *r*, series B3), *BFI* grows quickly, and then decays after the pumping is turned off. In the case of a slow energy input (small *r*, series B1) the *BFI* is smoothly increasing without significant jumps or drops, and remains constant when the energy input is stopped (Fig. 3a). Note that the eventual values of *BFI*s are different in these three series: fast input results in larger *BFI* at the end. Since the final states are characterized by the same intensity $\varepsilon$, larger *BFI* in case B3 should correspond to the narrower spectrum. Spectral shapes of the final states in experiments B1-3 are shown in Fig. 2, and in fact they look very much similar; the difference in their widths may be hardly seen by eye.

When the wave system receives energy, the skewness experiences somewhat larger fluctuations around zero, but it is still small. The difference in kurtosis, which characterizes the waves at the end of simulations, is clear for cases B1-3, see Fig. 3b. The system with smaller *BFI* (B1) has larger kurtosis *Ku*, what contradicts the common expectation supported by the theoretical relation obtained by Mori & Janssen (2006)

$$Ku = \frac{\pi}{\sqrt{3}} BFI^2 . \qquad (10)$$

According to (10) a larger *BFI* should lead to a large value of kurtosis (and, correspondingly, to higher probability of extreme waves). Formula (10) for *BFI* = 1 yields $Ku \approx 1.8$. While relation (10) is roughly correct for *BFI* and *Ku* before $T_{on}$, it is manifestly wrong for $t > T_{off}$. The strong deviation of the simulated irregular waves from a near-Gaussian stationary random process due to nonlinearity may be an evident explanation why formula (10) fails to describe the results.

The curves of probability distribution functions for wave amplitudes in Fig. 3c confirm the fact that the wave statistics deviates strongly from Gaussian. The curves for B1-3 correspond to the equilibrium state long after $T_{off}$; the Rayleigh distribution and the probability distribution before the stage of pumping ('initial state') are plotted as well. One may see that the wind effect further increases the probability of high waves, especially in cases B1 and B2.

The evolution of the probability in time is shown in Fig. 4 for cases B1 and B3. Isolines (see also the labels and the colour bar) display logarithmic units of the ratio of the calculated probability $P$ versus the Rayleigh distribution, $\log_{10}(P/P_{Rayleigh})$. One may see that the probability of amplitudes larger than $2\eta_{rms}$ grows shortly after start of simulations. Before the pumping is switched on, the probability of amplitudes twice exceeding $2\eta_{rms}$ is about $10^{1.5}$ times larger than prescribed by the Rayleigh law. The probability of high waves further increases after energy input, and it reaches $10^2$ times excess of the Rayleigh distribution (in the case B1, Fig. 4a). It is interesting to note the short period of even more extreme statistics just after $T_{off}$ in the case B3 (Fig. 4b). This peculiarity of the probability evolution is reflected by BFI (Fig. 3a, black curve) but is not captured by kurtosis (Fig. 3b).

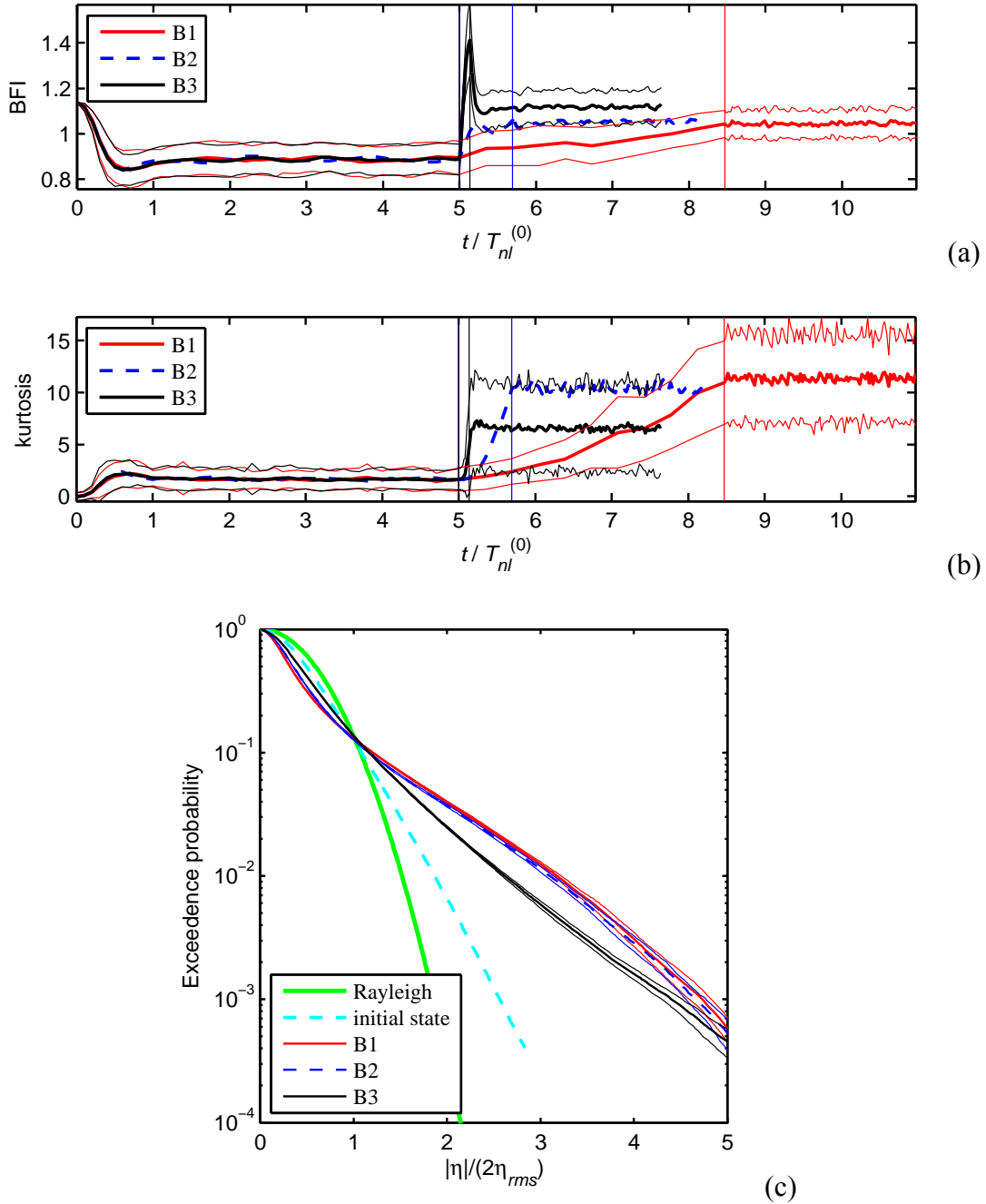

Fig. 3. The evolution of momentary *BFI* (a) and kurtosis (b) in three cases B1-3. The vertical lines mark the instants $T_{on}$ and $T_{off}$ for each of the experiments. The probability distribution

functions for the final equilibrium states compared with the Rayleigh distribution and the 'initial state' before forcing (c). Thin lines show the confidence intervals.

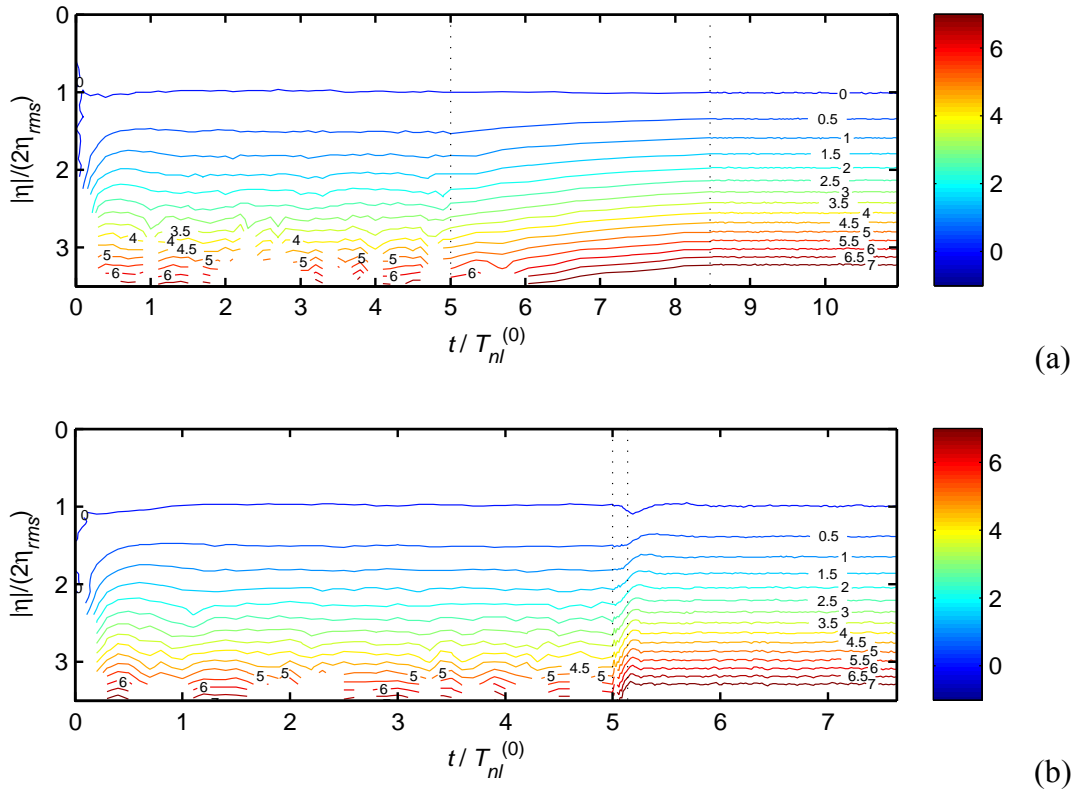

Fig. 4. Isolines with labels exhibit the difference $\log_{10}(P/P_{Rayleigh})$ between the exceedence probability function for wave amplitudes $P$ computed for simulated data, and the Rayleigh distribution $P_{Rayleigh}$, versus time for cases B1 (a) and B3 (b). The vertical dotted lines mark the instants $T_{on}$ and $T_{off}$ for each of the experiments.

Examples of the surface elevations are given in Fig. 5. They confirm that the simulation B1 with a very slow energy input results in more extreme wave groups than in the case B3. Comparing with the initial condition one may observe that waves at the end of the simulations form short and intense wave groups (the most intense wave groups in Fig. 5b,c are so narrow that include only 1-2 oscillations). The background small-amplitude waves are less discernible in case B1 (cf. Fig. 5b and Fig. 5c).

To conclude, the time-limited energy input results in distortion of the quasi-stationary state. Fast forcing (large $r$) leads to a rapid growth of *BFI* with further partial restoration; spikes of kurtosis are not observed, while momentary probability distributions testimony occurrence of extreme situation. The slow energy input results in a smooth monotonic growth of *BFI*, which eventually attains somewhat smaller values than in the case of a fast energy input. On the contrary, the new value of kurtosis, which characterizes the portion of high waves, is significantly larger in the case of adiabatically slow pumping. The later case eventually results in strongly non-Gaussian statistics with larger probability of high waves.

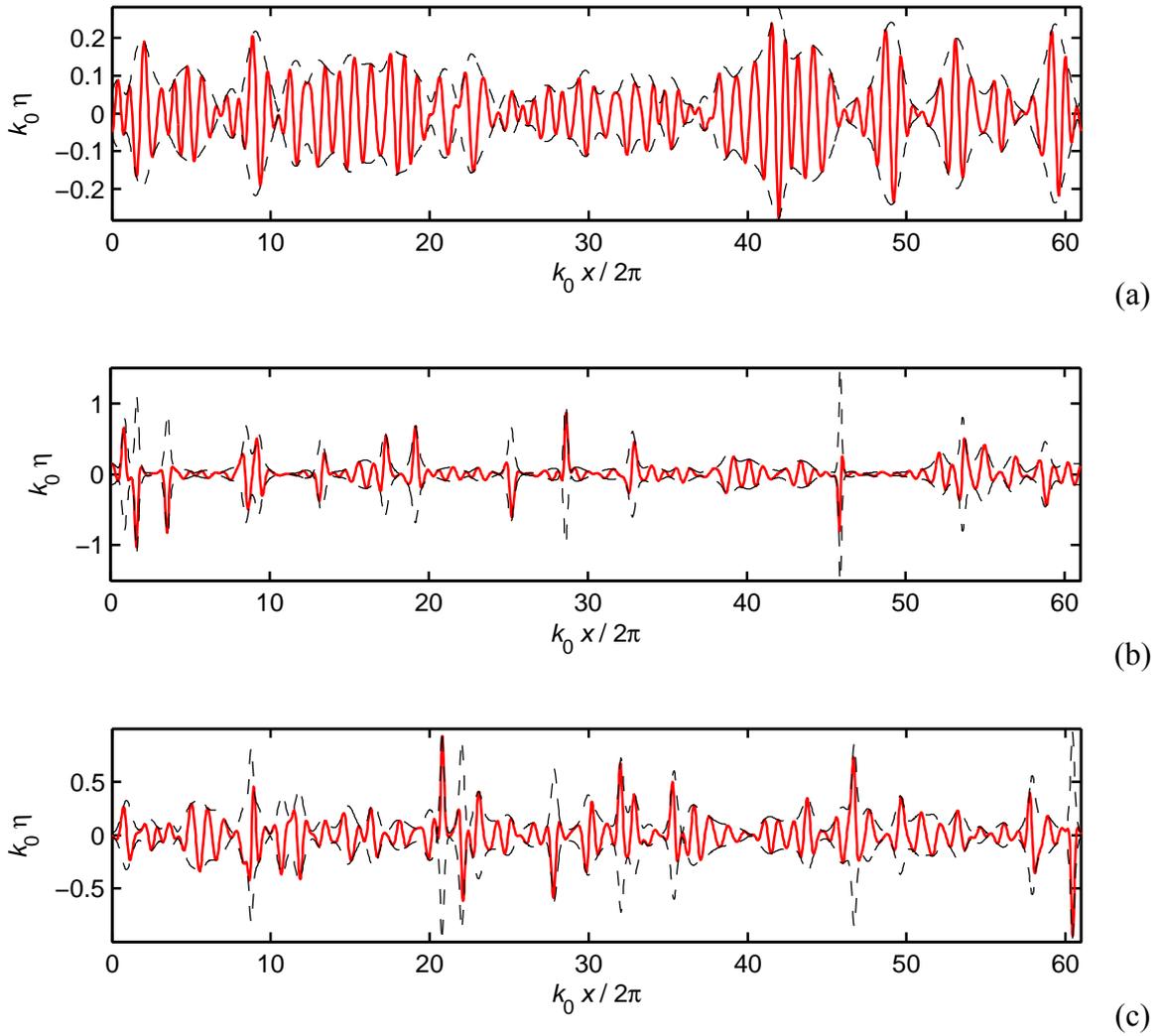

Fig. 5. Examples of surface elevations (solid lines) with envelopes $\pm k_0|A(x)|$ (dashed lines). The initial condition (a) and the final state (b) for case B1. The final state for case B3 (c).

## 4. Fast and adiabatic growth of solitary waves

It follows from Sec. 3 that the adiabatic pumping results in effective formation of intense and short soliton-like wave groups, which are readily seen in Fig. 5. The following numerical experiments (series S in Table 1) aim at consideration of a purified situation, when only one envelope soliton, the exact solution of the NLS equation, experiences external time-limited pumping. This problem was considered by Adcock and Taylor (2011) in case of an adiabatically slow change of an envelope soliton (numerically and with the help of approximate relations) and earlier by Kaup & Newell (1978) (in terms of the Inverse Scattering Technique) and Fabrikant (1984) (approximate description of the adiabatic change of the envelope soliton). Here we briefly consider different situations of fast and slow energy input from the point view of statistical description, and suggest the approximate description which qualitatively explains results observed in the numerical simulations.

The experiments are produced fully in line with simulations of driven irregular waves described in Sec. 3.2, though now the initial condition is taken in the form of an envelope soliton $A_s(x, t = 0)$, which in dimensional variables reads

$$A_s(x,t) = \frac{a_s}{\cosh\left(\sqrt{2}\,a_s k_0^2 x\right)} \exp\left[i\frac{(k_0 a_s)^2}{4}\omega_0 t\right]. \tag{11}$$

Here $a_s$ is a free parameter, which specifies the envelope soliton amplitude; the soliton width is inverse proportional to $a_s$ (therefore more intense groups are shorter, the dispersion is balanced by nonlinearity).

The dimensionless form (3) of the NLS equation may be written in this case as well, when the characteristic scales of the envelope soliton are used. Unlike irregular waves, envelope solitons represent the critical situation when waves are very much localized. The spectral width of envelope (11) may be calculated straightforwardly according to definition (9) (the Fourier spectrum for (11) may be found in Slunyaev et al, 2013): $v_s = (2/3)^{1/2} k_0 a_s$. The root-mean-square value of surface elevation in case of a localized solution (11) depends on the domain of consideration, and hence $\eta_{rms}$ is inappropriate characteristic of the wave intensity (for example, parameter of soliton density should be used in this case to avoid the ambiguity). Due to this reason the absolute values of other statistical moments, such as skewness and kurtosis (see (8)), are also meaningless. Anyway the characteristic steepness for an envelope soliton should be expressed in terms of $k_0 a_s$; if it is proportional to $k_0 a_s$, then all envelope solitons (11) are characterized by the same dimensionless value $BFI = O(1)$.

The parameters of pumping in simulations S1-3 (coefficients $G$, $r$ and the method how $T_{on}$ and $T_{off}$ are specified) are same as in corresponding experiments B1–3 (see Table 1). Experiment S1 corresponds to the slow energy input, and S3 – to the fast intense pumping. In contrast to series B, simulations S1-3 are performed for somewhat longer time after the pumping if discontinued – until $50 T_{nl}^{(1)}$.

When the forcing is slow, the envelope soliton grows in amplitude preserving its specific *sech* structure of the envelope soliton. When forcing is fast ($r$ is large), the envelope soliton transforms to a breathing intense wave group. The global maximum of the wave field after the stage of forcing is about $4a_s$ and is about the same in all the experiments (Fig. 6a). However, a crucial difference between the time dynamics in different cases may be observed in Fig. 6a. While in the regime of slow energy input (S1) the maximum is almost a constant for $t > T_{off}$, in the case of a rapid energy input the maximum oscillates, so that in case S3 it experiences about 100% variations. Naturally, the averaging over a few oscillations results in larger average amplitudes (and kurtosis) in the case of adiabatically slow input S1. It is evident that a similar result will take place in the case of ensemble of solitons with random initial phases $\omega_0 t$ in (11): adiabatically slow pumping will result in larger waves on the average. The Benjamin – Feir index is not monotonic (Fig. 6b); more interesting, it decreases in the situation of the most efficient energy transfer to the envelope soliton S1.

The spectrum of the initial condition (11) is shown in Fig. 7 by a green solid curve. Other curves show the spectra of waves after the stage of pumping. These spectra are computed by averaging over the long period $T_{off} + 25 T_{nl}^{(1)} < t < T_{off} + 50 T_{nl}^{(1)}$; they reveal the substantial difference between the fast and adiabatically slow pumping regimes: the former results in energy input into the spectral band of the carrier wave (S2, S3), while the latter one provides effective energy transfer from the domain of the carrier (S1). As a result, *BFI* drops down with time in the case S1, though the wave group grows in amplitude.

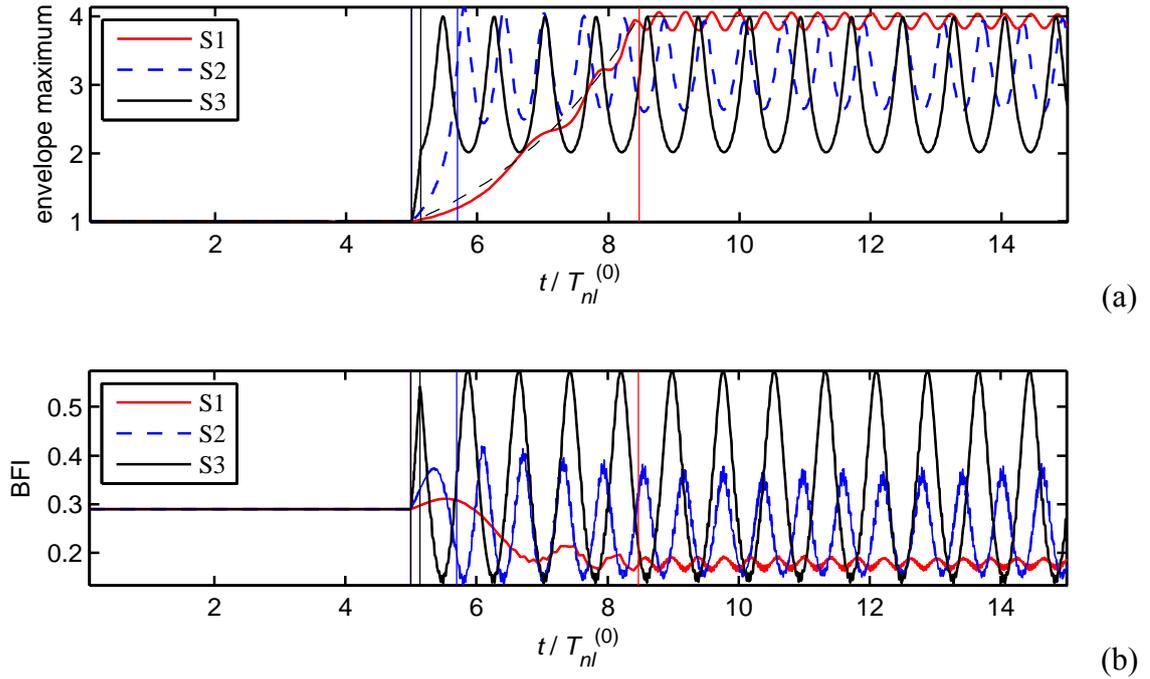

Fig. 6. Evolution of the envelope maximum (versus the initial value) (a) and BFI (b) in experiments S1-3. The vertical lines mark the instants $T_{on}$ and $T_{off}$ for each of the experiments. The thin broken line in panel (a) plots the adiabatic solution for case S1.

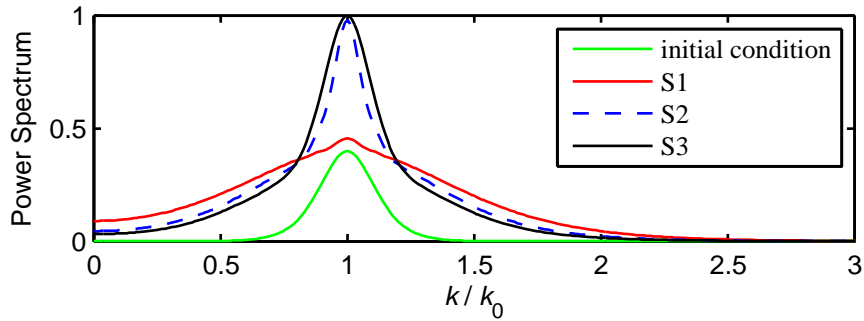

Fig. 7. Wavenumber power spectra of the surface elevation. The green solid curve shows the initial condition (the envelope soliton). Other lines show spectra in the quasi-stationary state after the pumping is turned off.

In the situation of adiabatically slow energy input the envelope soliton is preserving its structure, adjusting to the new condition (see e.g. Fabrikant (1984), Adcock & Taylor (2011)). Energy $E$ of the soliton is proportional to the product of squared soliton amplitude and soliton length. Since the latter is inverse proportional to the soliton amplitude, in the adiabatic regime the soliton amplitude is proportional to the energy. Consequently, amplification of the soliton amplitude in the adiabatic regime is $G^2$, what is equal to four in simulations of series S1-3. This dependence, predicted by theory, is shown by a thin broken line for experiment S1 in Fig. 6a. Meanwhile linear waves grow in time proportionally to $G$. The soliton evolves adiabatically when the characteristic nonlinear time is short in comparison with the characteristic forcing time; hence this regime requires parameter $r$ to be small. The characteristic time of dispersion for the envelope soliton is always in balance with the nonlinear time ($BFI \sim 1$).

In the limiting case of an abrupt wave pumping, the envelope obtains larger amplitude instantly, but the characteristic width of the envelope does not change (in this case the characteristic times of nonlinear adjustment as well as of dispersion are longer than the scale of forcing). As a result, the quickly amplified envelope soliton represents a pulse of the same width as initially, but with amplitude which is $G$ times larger than initially. This problem may be treated as the Cauchy problem for a humped disturbance within the NLS equation (in the same fashion as it was applied to the Korteweg – de Vries equation by Pelinovskii, 1971 and Johnson, 1973); this problem is solvable by means of the Inverse Scattering Transform (Zakharov & Shabat, 1972; Ablowitz et al, 1974). The solution to the scattering problem for *sech*-like pulse was found by Satsuma & Yajima (1974) and discussed in details in Kharif et al (2001). It follows that the amplified pulse will result in emergence of $N$ envelope solitons with amplitudes $A_n$, $n = 1, 2, …, N$, where

$$N = I\left[G + \frac{1}{2}\right], \qquad \frac{A_n}{A_s} = 2G + 1 - 2n. \qquad (12)$$

Here $I[\cdot]$ means the integer function. All the solitons, which emerge, are located at the same coordinate and hence belong to the kind of solutions called by Peregrine (1983) 'bi-solitons'. If no amplification happens $G = 1$, then $N = 1$ and $A_1 = A_s$. When $G > 1$, then the soliton grows in amplitude ($A_1 > A_s$). If $G > 3/2$, a new soliton appears. In experiments S1-3 we have $G = 2$; then formulas (12) give $N = 2$, and envelope soliton amplitudes are $A_1 = 3A_s$ and $A_2 = A_s$. The beating between solitons is clearly seen in Fig. 6a (case S3).

According to this simple analysis the general conclusion may be drawn. The maximum amplification of an envelope soliton is $G^2$ in the adiabatic regime, and is $2GN - N^2$ in the case of a rapid forcing. The latter estimate is obtained from (12), assuming that all the solitons may sum up at once; $N$ is the number of solitons, specified by (12). It is interesting to note that since $N \leq G$, then $2GN - N^2 \leq G^2$, – the adiabatic regime in general leads to a greater global maximum of the wave field, though the difference is small; these two amplification factors are equal when $G$ is integer. In simulations S1-3 the two regimes result in the same maximum envelope enhancement equal to four.

The transformation of the envelope soliton in the intermediate situation may be considered with the use of solution of the forced NLS equation discussed by Onorato & Proment (2012). They utilized smallness of the forcing term to obtain the approximate analytic solution of the forced equation basing on the exact analytic solution of the conservative NLS equation. According to Onorato & Proment (2012), the solution of the forced problem in dimensionless coordinates may be written in the form

$$A' \approx p\overline{A}'(px', pt')\exp\left[-i\frac{rp}{2}BFI^2 x'^2\right], \qquad p(t') = \frac{1}{1 - 2rt'}. \qquad (13)$$

Here $\overline{A}'(x', t')$ is a solution of the conservative dimensionless NLS equation (3) when $r = 0$. Relation (13) becomes trivial when $r = 0$. If $r > 0$, $p$ increases when time grows ($|rt'|$ should remain small).

Solution (13) consists of two parts. The product $pA_s'(px', pt')$ describes the envelope which in the course of evolution preserves the structure of the envelope soliton (i.e., its width is inverse proportional to the amplitude). The second part, the exponential factor, results in quadratic phase modulation. The typical scale of the modulation may be estimated as $2(\pi/pr)^{1/2}$. The ratio of the modulation length over the width of the envelope near the origin of the coordinate axis is much larger than the soliton width, when

$$\sqrt{\frac{\pi p}{r}} \gg 1. \qquad (14)$$

In these estimations we assume that for envelope solitons *BFI* ~ 1. If (14) is satisfied, then the modulation does not contribute to solution (13) significantly. Condition (14) is always satisfied when *r* is sufficiently small. If *r* is small, then the envelope soliton experiences the adiabatic growth, and no other waves are generated in the course of evolution.

If condition (14) is not fulfilled, the phase modulation in (13) cannot be neglected. The analysis of the inverse scattering problem for a pulse with quadratic modulation was performed by (under some approximations) Lewis (1985), Kaup & Malomed (1995), Slunyaev et al (2002). The quadratic 'chirp' changes amplitudes of the soliton part of the solution. According to the studies cited above, it may result in even further increase of the soliton amplitude (what qualitatively agrees with the consideration of an abrupt pumping given above).

It was not noticed by Onorato & Proment (2012) that the direct substitution of (13) into (3) yields

$$i\frac{\partial \overline{A}'}{\partial t'} + \frac{1}{BFI^2}\frac{\partial^2 \overline{A}'}{\partial x'^2} + |\overline{A}'|^2 \overline{A}' = -2i\overline{A}' r^2 t'(1 - 2rt'). \qquad (15)$$

To provide applicability of (13), the terms on the right should be small (then, the left-hand-side of (15) gives the conservative NLS equation), what requires sufficient smallness of $|rt'|$ = $|\Gamma t|$ ~ ln $G$ (when $t \sim T_w$). However, our numerical experiments show that the adiabatic regime may be accomplished rather well even when ln $G \approx 0.69$ (for $G = 2$).

**5. Discussion**

This research is aimed at investigation of the evolution of irregular nonlinear waves affected by external forcing. As a particular case, the target setting should model long-crested deep water waves driven by wind due to Miles' mechanism. We restrict our interest to only one very simplified model, the unidirectional nonlinear Schrödinger equation with linear forcing term. The ultimate benefit of this equation is the possibility of approximate transformation of its solutions to the framework of the integrable version of the NLS equation; thus analytic description of non-trivial effects becomes available. Though the original physical problem involves many parameters, the eventual scaled equation (3) depends on only two key parameters.

For magnification of the physical effects for illustration purposes, we consider rather exaggerated regimes so that the amplified groups are narrow, and the sea surface waves (Fig. 4b,c) obviously would break in reality. On the other hand, recently solutions of the NLS equation have been found describing the strongly nonlinear dynamics of water waves sometimes surprisingly well. We will mention here just a few examples: i) the NLS envelope soliton solution was used to generate almost breaking steep solitary wave groups (with only a few oscillations within the group) in laboratory and numerical simulations by Slunyaev et al (2013); ii) the NLS rational multi-breather solutions were shown to be replicable in experimental conditions for rather steep waves (see Chabchoub et al (2011) and the sequel of works by this team); iii) the NLS model is capable of adequate description of statistics of steep nonlinear wave groups evolution in a flume (Shemer et al, 2010b). Thus, we believe that the discussed effects in wave dynamics may be observed in reality with some reasonable alteration. For example, the beating of solitons (see discussion of Fig. 6, case S3) can change to fission due to effects of non-integrability (Chabchoub et al, 2013b). The effect of directionality, when coherent solitary groups disintegrate, is probably the biggest challenge to the effects considered in our study.

Different regimes are studied depending on the ratio between characteristic time scales of the energy input and of the adjustment due to nonlinearity. In addition, the external forcing is limited in time, what allows consideration of the wind-driven system evolution from one quasi-stationary state to another. The non-stationary evolution of nonlinear waves is prone to extreme statistics, and we focus on extreme waves with BFI slightly exceeding the threshold $BFI = 1$ through tracking the evolution of the fourth statistical moment, kurtosis, the compound spectral parameter, the Benjamin – Feir Index, and also the wave amplitude probability distribution.

The time-limited forcing of irregular waves breaks the statistical equilibrium; the existence (and emergence) of coherent nonlinear wave groups results in strong deviation from the Gaussian statistics, what becomes evident due to, in particular, large values of kurtosis. Effects of nonlinear adjustment and external pumping on irregular wave ensembles are studied with the help of stochastic numerical simulations; they build up the rich picture of nonlinear wave dynamics. Auxiliary deterministic simulations of envelope solitons are exploited to clarify the details of evolution of wave groups which are constrained by nonlinearity.

In the general case irregular waves governed by the NLS equation are represented by two parts which correspond to different kinds of the spectrum of the associated scattering problem: envelope solitons (discrete spectrum) and dispersive quasilinear waves (continuous spectrum). The solitons are nonlinear coherent wave groups which within the framework of the integrable NLS equation preserve their structure and interact with other waves elastically. They have fundamental significance in solving the Cauchy problem for the NLS equation and in interpretation of the picture of instability of modulated waves.

Our investigation shows that very different scenarios of irregular wave amplification take place depending on parameter $r$ which is proportional to the ratio of characteristic nonlinear time versus characteristic time scale of the forcing. For the given value of wave amplitude enhancement $G$, the adiabatically slow pumping of the wave system leads to the increase of envelope soliton amplitudes proportional to $G^2$. Consequently, the portion of solitary waves compared to quasi-linear waves is increasing, kurtosis grows and the portion of high waves grows as well. The wave spectrum becomes wider due to arising heavy tails. This result agrees with Adcock and Taylor (2011), who state that "energy input enhances any soliton-like structures relative to regular waves".

In contrast, the abrupt forcing results in increase of envelope soliton amplitudes proportional to $G$, but new solitons may emerge. Due to accordant interactions, the solitons may generate waves with amplitudes $\sim G^2$ (similar to the adiabatic regime), but in average the quickly amplified nonlinear waves are characterized by a smaller portion of very high amplitudes (smaller kurtosis) and a narrower spectrum. At the same time, when the instantaneous forcing is strong, large rapid surges of the Benjamin – Feir Index may be observed associated with increase of large wave probability, which relax within time less than $T_{nl}$.

So far, the Benjamin – Feir Index seems to be the only parameter characterizing the sea state, which is an intelligible candidate for the rogue wave warning criteria. Conditions with $BFI > 1$ favour the modulational instability which causes rogue waves. Unfortunately, its application to the real forecast is not straightforward due to at least several reasons (here we confine our interest to the situation of long-crested waves): i) the method how the $BFI$ is calculated should be robust; ii) the spread of BFIs calculated for realistic seas is not large, typically *BFI* ~ 1; thus *BFI* should be calculated with high accuracy to be conclusive; iii) there is a memory effect: the instants with large *BFI* and extreme wave events may be spased within a few $T_{nl}$, what for realistic conditions gives the time slot of tens minutes.

We emphasize that even in our purified computer simulations some weak modification of the approach to calculate BFI may result in very different estimations of its value. This fact is illustrated with help of Fig. 8. The figure reproduces the evolution of *BFI* with time in experiments B1-3 (similar to Fig. 3a), but the width of spectrum (9) is calculated on the basis of the entire domain of wavenumbers $k \geq 0$. Though within the time interval $t < T_{on}$ there is no noticeable difference between Fig. 3a for *BFI*($t$) and Fig. 8 for the modified index $BFI^*(t)$, the final values of *BFI* and $BFI^*$ differ substantially. While the values of *BFI* got increased after the energy input, $BFI^*$ decrease in all the cases.

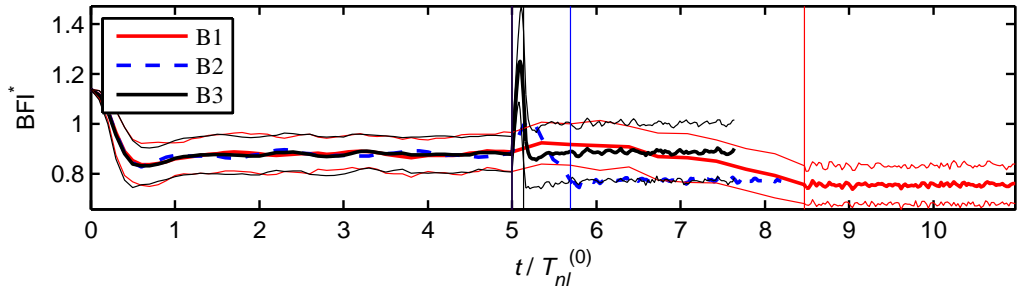

Fig. 8. Same as Fig. 3a, but the spectrum width which enters the BFI (5) is calculated on the basis of the entire domain of wavenumbers $k \geq 0$ in (9).

The Benjamin – Feir Index which appears in the NLS framework (Onorato et al, 2001; Janssen, 2013) is now considered to be the main possible candidate to the role of a rogue wave indicator (despite significant criticisms, see e.g. review in Kharif et al, 2009 and discussion above). Of course, for application purposes the index should be modified to take into account the effect of wave directionality (e.g. Mori et al, 2011; Ruban et al, 2011). The explicit relation between wave kurtosis and BFI was obtained in Mori & Janssen (2006), which for dynamical kurtosis gives (10), though their theory is based on assumptions which often fail. The cancellation of the theory by Mori & Janssen (2006) in strongly non-stationary evolution of steep water waves was shown in Slunyaev & Sergeeva (2011) (see their Fig. 3); the present work gives other examples, when large kurtosis corresponds to small *BFI*. Having a look at Figs. 3b, 4, 5, it becomes obvious that we deal with strongly non-Gaussian dynamics, when traditional statistical theories are not applicable.

The NLS equation is a general model which is popular in many physical branches; it provides a brilliant balance between physical conformity and mathematical accessibility for many applications. Extreme or rogue waves in the sea, optics and other fields are to the leading order described by the NLS equation (see Onorato et al, 2013), hence our results have obvious application to phenomena in other media. Long-living soliton-like wave groups on the water surface may be observed in the conditions of long-crested waves (Slunyaev et al, 2013), or when trapped by bathymetry (Pelinovsky et al, 2010) or by currents (Shrira & Slunyaev, 2014). We may expect that the conclusions which we derive basing on the NLS framework will remain valid for other systems which support long-living soliton-like wave patterns: adiabatic energy pumping will result in selective amplification of the soliton-like waves in a greater extent than quasi-linear waves (in accordance with Adcock &Taylor, 2011). This should be an efficient mechanism for realization of conditions for strongly non-Gaussian wave statistics.

In the situation of integrable or near integrable wave systems, the 'nonlinear' spectrum of the associated scattering problem could provide clues for understanding of the essentially nonlinear wave dynamics and even elaboration of warning criteria in situations when the risk of unacceptably large waves is high. The relation between the 'nonlinear' spectrum and the wave statistics represents an important and complicated problem, which is partly addressed by our study.

The evolution of irregular waves over non-uniform bed was studied numerically (variable-depth Korteweg – de Vries equation and NLS equation) and experimentally by Sergeeva et al (2011) and Trulsen et al (2012), Zeng & Trulsen (2012). They observed situations when shoaling irregular waves exhibited unexpectedly extreme statistics. It was pointed out that the transition stage from deep to shallow water was a necessary but not sufficient condition for occurrence of an extreme state. We anticipate that the extreme dynamics of shoaling waves may be observed when the condition of adiabatic evolution of soliton waves is fulfilled; though this hypothesis requires further investigation.

**Acknowledgements**

The authors were supported by RFBR grants 14-02-00983 and 14-05-00092, and also State Contract 2014/133 (project 2839). The support from Volkswagen Foundation is gratefully acknowledged.